\documentclass[a4paper,12pt]{article}
\usepackage{graphicx}
\usepackage{subfigure}
\usepackage{amsmath}
\usepackage{amssymb}
\usepackage{float}
\usepackage[margin=0.8in]{geometry}

\begin{document}

\title{Charge asymmetry dependency of $\pi^{+}/\pi^{-}$ elliptic flow in Au + Au collisions at $\sqrt{s_{NN}}$ = 200 GeV}

\author{Hongwei Ke$^{1, 2}$ (for the STAR Collaboration) \\
\small{$^1$ Institute of Particle Physics, Central China Normal University, Wuhan 430079, China} \\
\small{$^2$ Brookhaven National Laboratory, Upton, NY 11973, USA}}
\date{}
\maketitle

\begin{abstract}
In this proceedings, we present STAR's measurement of $v_{2}$ difference between positively charged pions and negatively charged pions at low transverse momentum for Au + Au collisions at $\sqrt{s_{NN}}$ = 200 GeV. The integrated $v_2$ of $\pi^+$ ($\pi^-$), $0.15 < p_T < 0.5 \text{ GeV}/c$, decreases (increases) linearly with the increasing charge asymmetry $A_{\pm}$. The $v_2$ difference between $\pi^+$ and $\pi^-$ is proportional to $A_{\pm}$ and the slope parameters have their order of magnitude of about 0.01. All of these observed features are consistent with the predication based on Chiral Magnetic Wave. The centrality dependence of the $v_2$ difference is different than that was predicted.
\end{abstract}

\section{Introduction}
In relativistic heavy ion collisions, a significant fraction of incident energy is deposited in the collision region. Due to the existence of the large energy density, a phase of extremely hot and dense matter consisting of quarks and gluons, the Quark Gluon Plasma (QGP) \cite{Arsene2005, Back2005, Adams2005, Adcox2005}, is expected to be created. On the other hand, in non-central collisions, spectators keep moving along the incident directions. Since spectators are charged objects, moving spectators will generate electric current which induces a magnetic field perpendicular to the reaction plane (defined by the line connecting the colliding nuclei and the incident direction).  It has been estimated that the magnetic field created in heavy ion collisions can be as strong as up to $\sim 10^{15}$ Tesla at the beginning of collisions while it decreases rapidly with time \cite{Kharzeev2008}.
With QGP under a such strong magnetic field, it is pointed out that the chiral magnetic effect (CME) \cite{Kharzeev2008, Dmitri2006, Kharzeev2007, Fukushima2008, Kharzeev2010} and the chiral separation effect (CSE) \cite{Son2004, Metlitski2005} are expected to exist. 
The CME in Quantum Chromodynamics (QCD) is that the external magnetic field induces a vector current with a finite axial chemical potential $\mu_A$, which describes the asymmetry between the densities of left- and right-handed quarks.
\begin{equation}
\label{eq:cme}
\boldsymbol{j}_V = \frac{N_c e}{2 \pi^2} \mu_A \boldsymbol{B}.
\end{equation}
If there exist meta-stable {\it P} and {\it CP} odd domains, space-time regions occupied by a classical field with a nonzero topological charge, under CME it would lead to the asymmetry in the emission of positively and negatively charged particles perpendicular to the reaction plane in non-central collisions. Such effect has been intensively studied by several experiments \cite{Abelev2009b, Abelev2010b, Ajitanand2010, Selyuzhenkov2012}. The other effect in the discussion, CSE, is the separation of chiral charge along the axis of the external magnetic field at finite density of vector charge. The resulting axial current is given by
\begin{equation}
\label{eq:cse}
\boldsymbol{j}_A = \frac{N_c e}{2 \pi^2} \mu_V \boldsymbol{B}.
\end{equation}
The local fluctuation of axial current from Eq. (\ref{eq:cse}) will induce a local fluctuation of the axial chemical potential, and thus according to Eq. (\ref{eq:cme}) a fluctuation of electric current. The resulting fluctuation of electric charge density will then in turn, again, induce an axial current according to Eq. (\ref{eq:cse}). With alternating CME and CSE this cycle continues, and it gives rise to a gapless excitation regarded as Chiral Magnetic Wave (CMW). Please note that the CMW does not require the local parity violation although one of its ingredients, namely the CME, is responsible for making the local parity violation signal visible in relativistic heavy ion collisions.

The CMW has a consequence that also could be observed in experiments. As pointed out in a recent theoretical work \cite{Burnier2011a}, a CMW in QGP will form a electrical quadrupole moment which can lead to more positive charge near the poles of the created fireball and more negative charge near the equator. That means, this configuration will cause a difference in elliptic flow, $v_2$ \cite{Poskanzer1998}, between positively and negatively charged particles. Taking into account the quadrupole moment, the azimuthal angle distribution of positive and negative particles is given by Eq. (\ref{eq:azimuthalDist}),

\begin{equation}
\begin{split}
\label{eq:azimuthalDist}
   \frac{d N_{\pm}}{d \phi} & = N_{\pm} \left[ 1 + 2 v_2 \cos(2 \phi) \right] \\
       & \approx \bar{N}_{\pm} \left[ 1 + 2 v_2 \cos(2 \phi) \mp A_{\pm} r \cos(2 \phi) \right]
\end{split}
\end{equation}
Here $\bar{N}_+ \:(\bar{N}_-)$ is number of positive (negative) particles, $A_{\pm} \equiv (\bar{N}_+ - \bar{N}_-)/(\bar{N}_+ + \bar{N}_-)$ is the net charge asymmetry and $r \equiv 2 q_e / \bar{\rho_e}$, while $q_e$ represents the electrical quadrupole and $\bar{\rho_e}$ is charge density. Authors in \cite{Burnier2011a} also pointed out this difference in elliptic flow should be see via $\pi^{+}$ and $\pi^{-}$ due to their similar absorption cross sections in hadronic matter at finite baryon density. The elliptic flow difference between $\pi^{+}$ and $\pi^{-}$ is expected to be proportional to the net charge asymmetry as described in Eq. (\ref{eq:v2Diff}).

\begin{equation}
\label{eq:v2Diff}
\Delta v_{2}^{\mathrm{CMW}} \equiv v_{2}(\pi^-) - v_{2}(\pi^+) \approx r A_{\pm}
\end{equation}

There are other models that also predicted different elliptic flow between particles and anti-particles. Quark transport model predicted $v_2$ order of particles and anti-particles according to their quark components \cite{Dunlop2011}. This model is base on quark coalesce and assumes transported quarks have lager $v_2$ than produced quarks. AMPT model with hadronic potential also predicts different $v_2$ of particles and anti-particles \cite{Xu2012a}. However, as currently implemented the two models mentioned above did not provide a net charge asymmetry dependency of the $v_2$ difference.
	
In STAR experiment, the difference in integrated $v_2$ between particles and anti-particles has been observed. In this paper, we aim to study such difference differentially as a function of net charge asymmetry, and compare that to features predicted by the CMW.

\section{Data Analysis and Results}
STAR detector complex \cite{Ackermann2003} has large acceptance and full azimuthal coverage, which is ideal for studying elliptic flow. In this analysis, we used $\sim$ 238M minimum-bias Au + Au events at $\sqrt{s_{NN}} = $ 200 GeV taken by STAR during year 2010. STAR Time Projection Chamber (TPC) \cite{Anderson2003} is used to reconstruct tracks and identify particles. All particles used in this analysis are required to be in the middle rapidity range $|\eta| < 1.0$. For flow analysis, pions are identified by ionization energy loss in TPC and required to have $0.15 < p_{T} < 0.5 \text{ GeV}/c$. All charged particles within $0.15 < p_{T} < 12 \text{ GeV}/c$ are used to calculate net charge asymmetry except low $p_T$ protons, $p_{T} < 0.4 \text{ GeV}/c$. Those protons are dominated by protons produced from knockout/nuclear interactions of pions with inner detector material, and thus rejected. Antiprotons within the same $p_T$ range are also excluded from the net charge asymmetry calculation in order to balance the cut on protons.

To estimate the elliptic flow of $\pi^+$ and $\pi^-$, we use Q-cumulants method, which is also called direct cumulants method \cite{Bilandzic2010}. Q-cumulants method estimates different order of harmonics via multi-particle correlations. This relatively new method needs only on pass over tracks and has comprehensive detector inefficiency corrections. In practice of this analysis, two-particle correlation is used to estimate $v_{2}\{2\}$ of pions and an $\eta$-gap of 0.3 pseudo-rapidity unit on each side is applied to suppress the short-range correlation.

\begin{figure}[htbp]
\begin{center}
\includegraphics[width=4in]{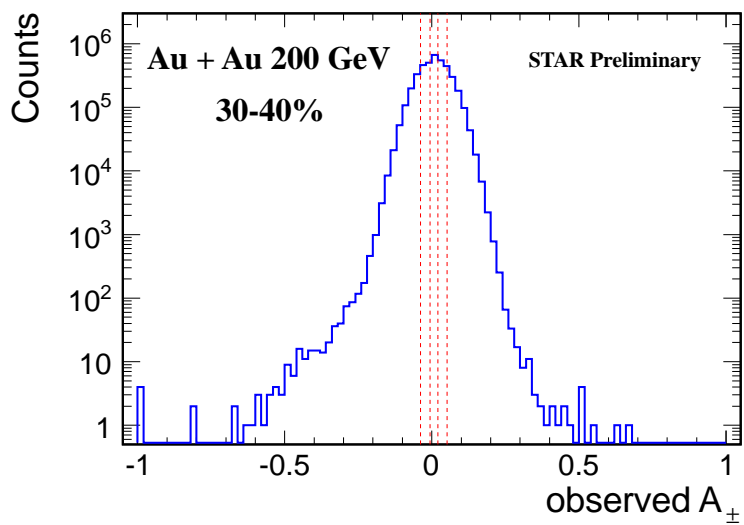}
\caption{Observed net charge asymmetry $A_{\pm}$ distribution of 30-40\% most central Au + Au events at $\sqrt{s_{NN}} = 200 \text{ GeV}$. The vertical dashed red lines show the sub-groups of events.}
\label{fig:chAsym}
\end{center}
\end{figure}

With in a certain centrality bin, we calculate the net charge asymmetry $A_{\pm}$ event by event and divide the sample into five sub-groups so that each sub-group has roughly the same number of events. Considering finite detector acceptance and tracking inefficiency, $A_{\pm}$ obtained directly from final state hadrons is different than the true $A_{\pm}$, thus it needs to be corrected.  In discussion below, we call the $A_{\pm}$ before the correction the observed $A_{\pm}$, and otherwise, the true  $A_{\pm}$. The correction procedure on the observed $A_{\pm}$ will be discussed later. Fig. \ref{fig:chAsym} shows the observed $A_{\pm}$ distribution of 30-40\% most central Au + Au events at $\sqrt{s_{NN}} = 200 \text{ GeV}$, in which red lines divide the statistics equally into five bins.

\begin{figure} 
\begin{center}
\subfigure{\label{fig:v22} 
\includegraphics[width=3.1in]{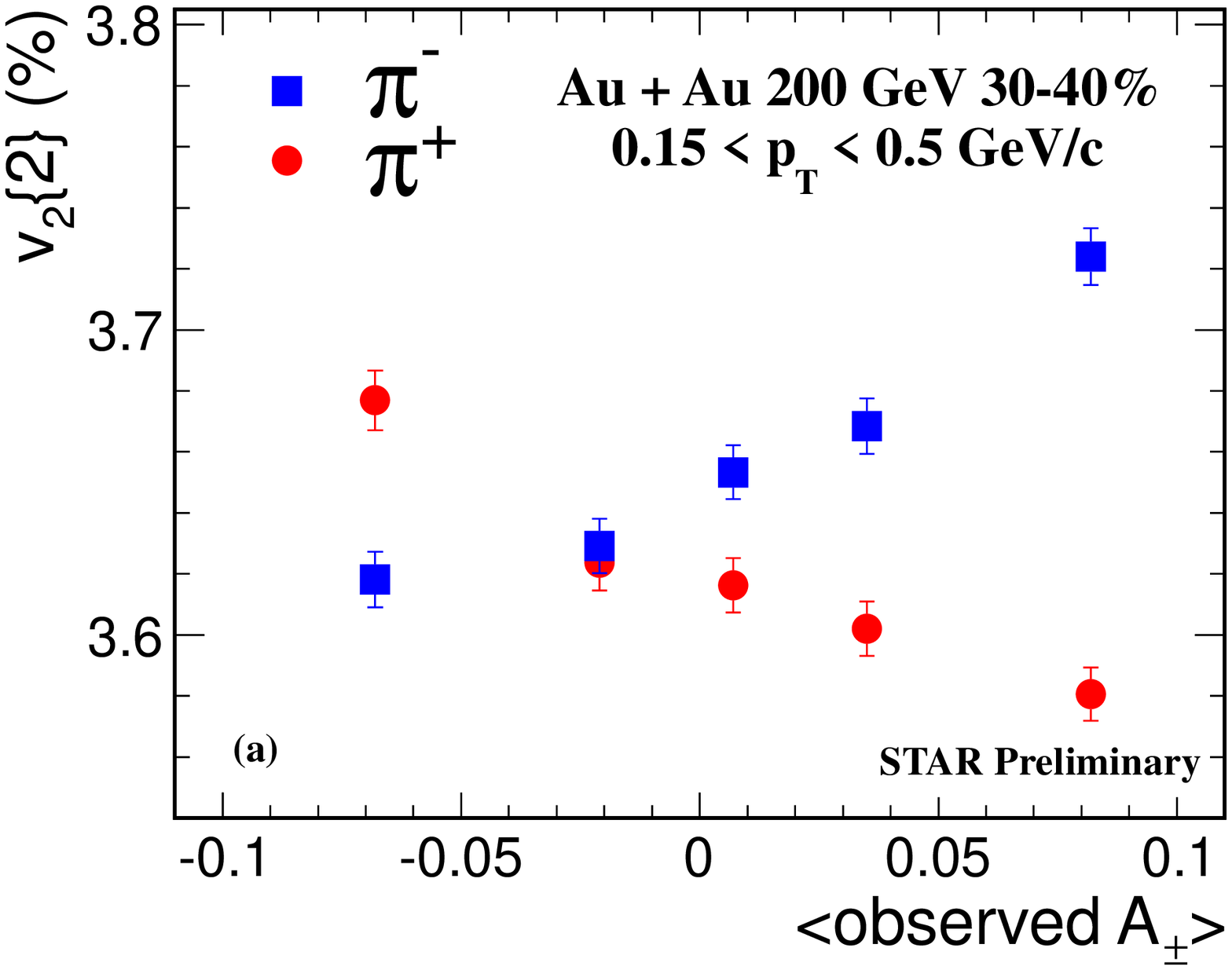} }
\subfigure{ \label{fig:v22_fit}
\includegraphics[width=3.1in]{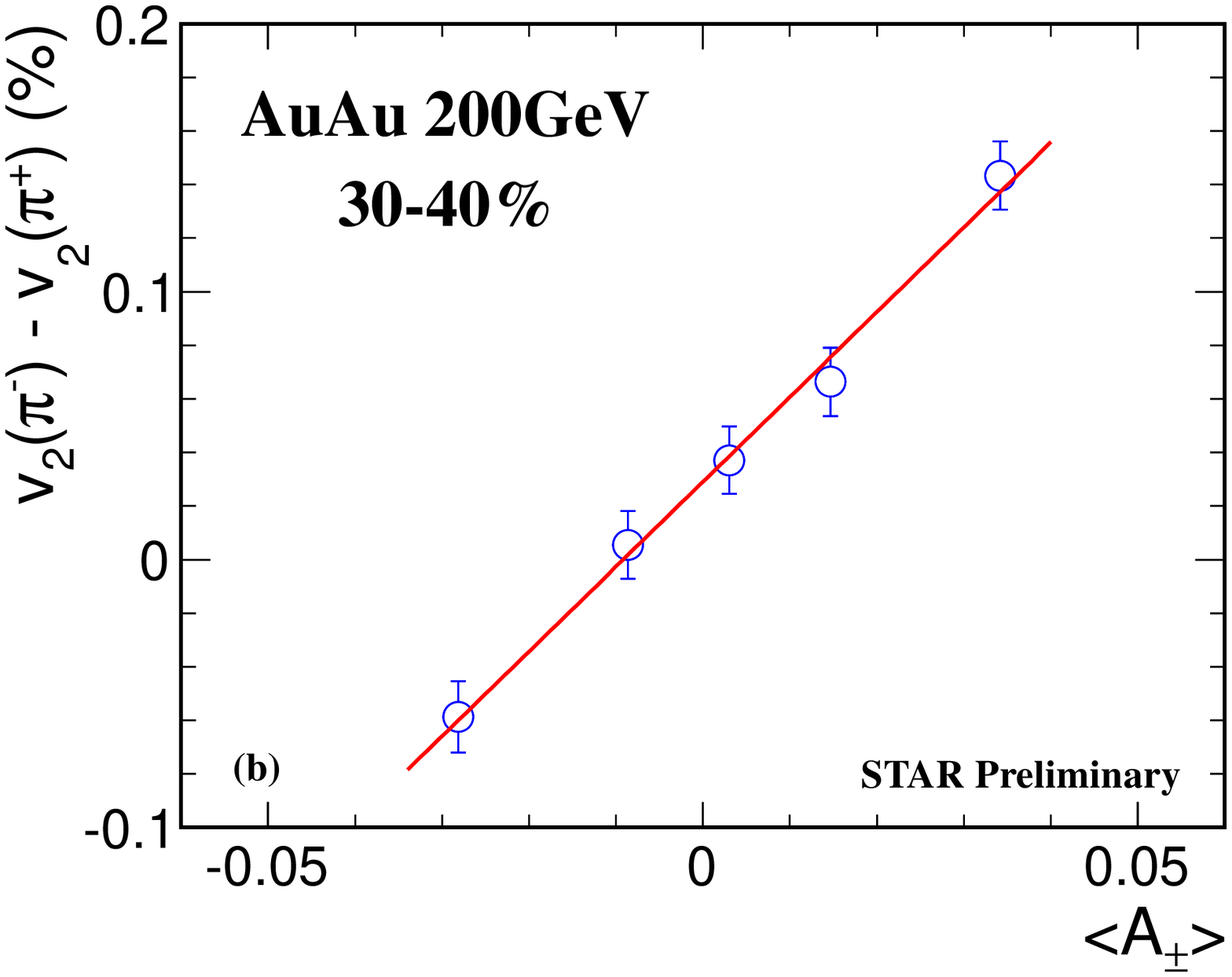} }
\caption{(a) Integrated $v_2\{2\}$ of $\pi^+$ and $\pi^-$, $0.15 < p_{T} < 0.5 \text{ GeV}/c$, shows as a function of observed $A_{\pm}$, 30-40\% most central Au + Au events at $\sqrt{s_{NN}}=200\text{ GeV}/c$. (b) $v_2$ difference with respect to corrected net charge asymmetry $A_{\pm}$ fits to a straight line.} 
\end{center}
\end{figure}

We measured the integrated $v_{2}\{2\}$ of $\pi^+$ and $\pi^-$ with $0.15 < p_{T} < 0.5 \text{ GeV}/c$ and present it as a function of observed $A_{\pm}$. In Fig. \ref{fig:v22}, $v_{2}$ of $\pi^+$ and $\pi^-$ both show a linear relationship with respect to the observed $A_{\pm}$. With the increasing observed $A_{\pm}$, $v_{2}$ of $\pi^-$ increases and $v_{2}$ of $\pi^+$ decreases. As a consequence, the $v_2$ difference between $\pi^+$ and $\pi^-$, $\Delta v_{2} = v_{2}(\pi^-) - v_{2}(\pi^+)$, increases linearly with the increase of the observed $A_{\pm}$. All of the features mentioned here are consistent with the prediction made in \cite{Burnier2011a}.  

The slope of the linear relationship between $\Delta v_{2}$ and $A_{\pm}$ is of special interest because it has a connection to the electrical quadrupole induced by a CMW in QGP. However, one needs to correct the observed $A_{\pm}$ before a meaningful slope can be extracted. Applying the correction on observed $A_{\pm}$ is conducted in two steps. First, we measure the tracking efficiency of charged pions in embedding data and use it as the tracking efficiency of all charged hadrons. This efficiency reflects the combined effect of detector and track reconstruction inefficiency. Secondly, we generated some Monte-Carlo events by using the event generator Hijing \cite{Gyulassy1994}. Net charge asymmetry in Hijing events is measured before and after applying the tracking efficiency and all of track selection cuts as applied on real data. The relationship between the two measurements of net charge asymmetry is used to map the observed $A_{\pm}$ to the true $A_{\pm}$. Roughly, the corrected charge asymmetry $A_{\pm}$ is a half of its corresponding observed $A_{\pm}$.

In Fig. \ref{fig:v22_fit}, the $\Delta v_{2}$ is plotted a function of $A_{\pm}$. A clearly linear relationship is observed. Fitting the data points to a straight line, the slope parameter can be extracted. This procedure is repeated for all centrality classes.

\begin{figure}[H]
\begin{center}
\subfigure{ \label{fig:slopes} 
\includegraphics[width=3.1in]{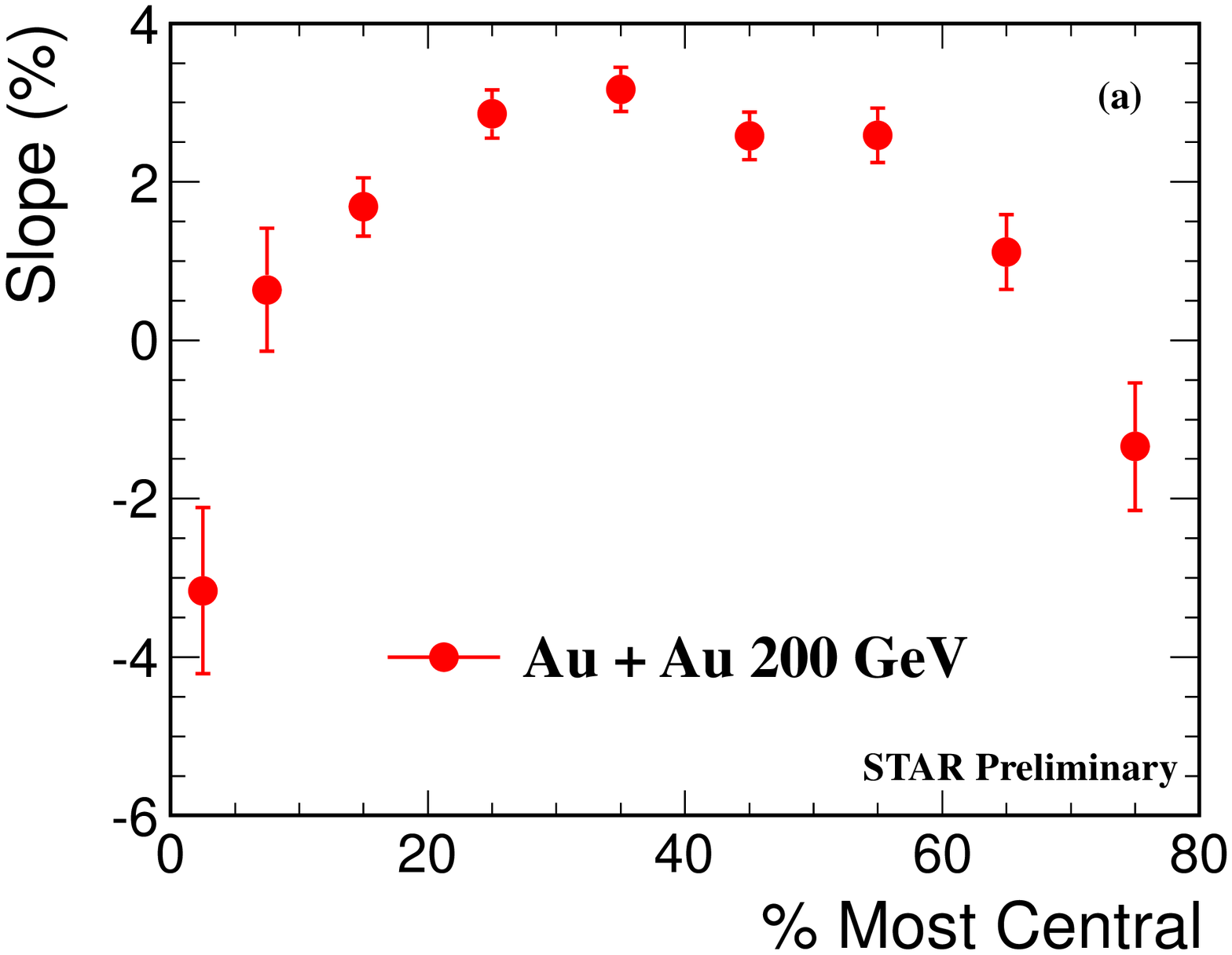} }
\subfigure{ \label{fig:slopes_comp}
\includegraphics[width=3.1in]{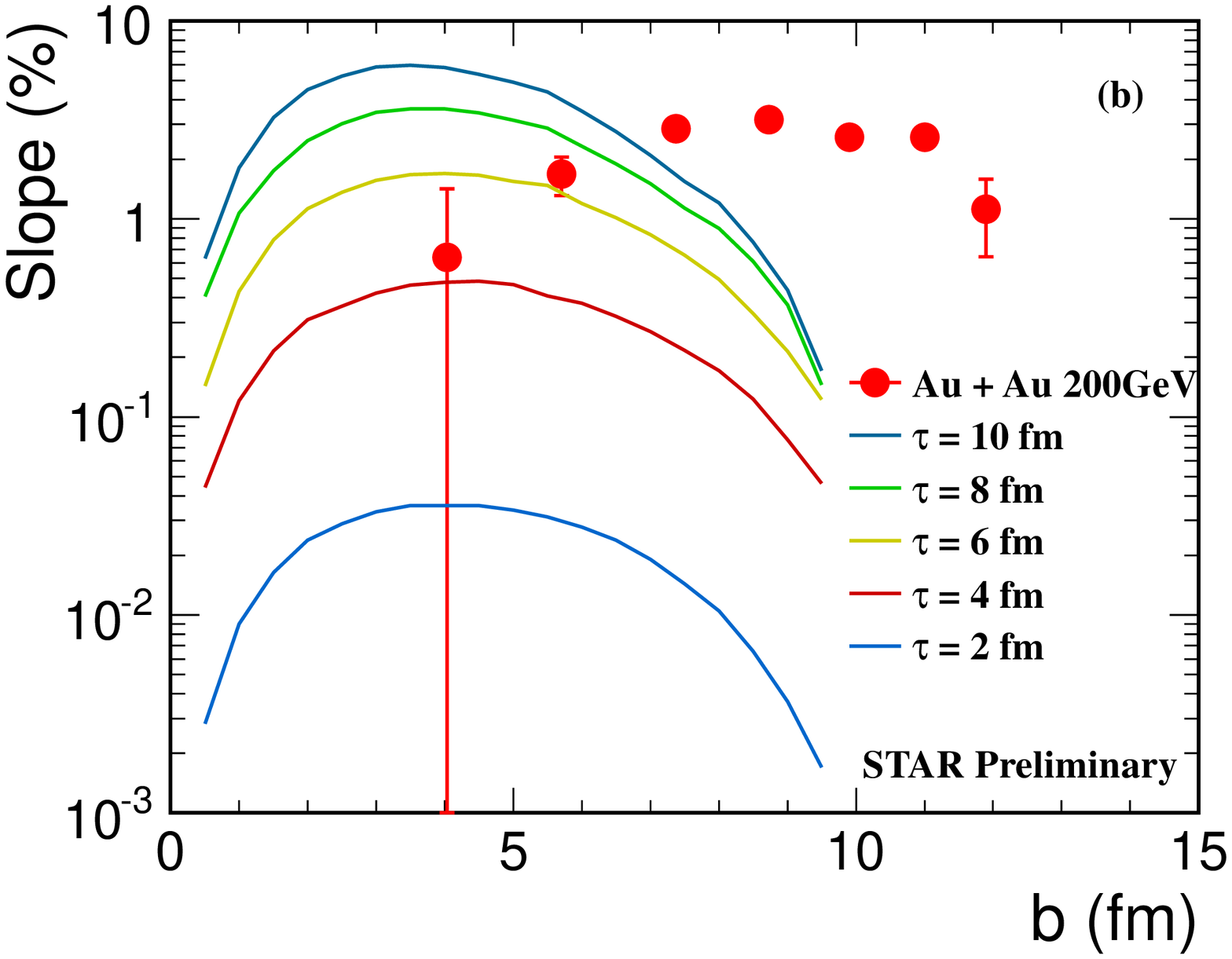} }
\caption{(a) Slope extracted from the relationship of $\Delta v_2$ with respect to $A_{\pm}$ for all centralities. Statistical errors only. (b) Slope parameters compare to the theoretical prediction made in \cite{Burnier2011a}, which is shown by the solid lines.} 
\end{center}
\end{figure}

The slope parameters measured for events in the nine centrality bins are shown in Fig. \ref{fig:slopes}. Here only statistical uncertainties are shown. It can be seen that slopes have the order of magnitude of about $0.01$, which is consistent with the theoretical prediction based on CMW. The slopes are relatively large in mid-centrality and relatively small in peripheral and central collisions. In order to quantitatively compare to the prediction made in \cite{Burnier2011a}, we map out our centrality definition to impact parameter $b$ according to a previous STAR study based on Monte-Carlo simulation \cite{Abelev2009}. The comparison is shown in Fig. \ref{fig:slopes_comp}. It can be seen that, although the magnitude of slopes in general agree with the prediction, the centrality dependency is a little different - the data is peaked at a larger $b$ than the prediction. As the strong correlation exist between $b$ and the magnetic field, this observation calls for a more advanced calculation on the magnetic field that is used to calculate the $\Delta v_2$.

\section{Summary}
We measured the $v_2$ of $\pi^+$ and $\pi^-$ of Au + Au collisions at $\sqrt{s_{NN}} = 200$ GeV in different centrality and different net charge asymmetry $A_{\pm}$ bins. The integrated $v_2$ of $\pi^+$ ($\pi^-$), $0.15 < p_T < 0.5 \text{ GeV}/c$, decreases (increases) linearly with the increasing $A_{\pm}$. The $v_2$ difference between $\pi^+$ and $\pi^-$ is proportional to $A_{\pm}$ and the slope parameters have their order of magnitude of about 0.01. All of these observed features are consistent with the predication based on CMW in \cite{Burnier2011a}. However, the predicted trend of the slope with respect to impact parameter is different than the measurement.

\section{Acknowledgements}
This work is supported in part by National Natural Science Foundation of China under Grants 11075060, 11135011 and 11105060.

\bibliographystyle{iopart-num}
\bibliography{paperLib}

\providecommand{\newblock}{}
\begin{thebibliography}{10}
\expandafter\ifx\csname url\endcsname\relax
  \def\url#1{{\tt #1}}\fi
\expandafter\ifx\csname urlprefix\endcsname\relax\def\urlprefix{URL }\fi
\providecommand{\eprint}[2][]{\url{#2}}

\bibitem{Arsene2005}
Arsene I and et~al (BRAHMS Collaboration) 2005 {\em Nuclear Physics A\/} {\bf
  757} 1 -- 27

\bibitem{Back2005}
Back B and et~al (PHOBOS Collaboration) 2005 {\em Nuclear Physics A\/} {\bf
  757} 28 -- 101

\bibitem{Adams2005}
Adams J and et~al (STAR Collaboration) 2005 {\em Nuclear Physics A\/} {\bf 757}
  102 -- 183

\bibitem{Adcox2005}
Adcox K and et~al (PHENIX Collaboration) 2005 {\em Nuclear Physics A\/} {\bf
  757} 184 -- 283

\bibitem{Kharzeev2008}
Kharzeev D~E, McLerran L~D and Warringa H~J 2008 {\em Nuclear Physics A\/} {\bf
  803} 227 -- 253

\bibitem{Dmitri2006}
Dmitri and Kharzeev 2006 {\em Physics Letters B\/} {\bf 633} 260 -- 264

\bibitem{Kharzeev2007}
Kharzeev D and Zhitnitsky A 2007 {\em Nuclear Physics A\/} {\bf 797} 67 -- 79

\bibitem{Fukushima2008}
Fukushima K, Kharzeev D~E and Warringa H~J 2008 {\em Phys. Rev. D\/} {\bf
  78}(7) 074033

\bibitem{Kharzeev2010}
Kharzeev D~E 2010 {\em Annals of Physics\/} {\bf 325} 205 -- 218

\bibitem{Son2004}
Son D~\ T and Zhitnitsky A~R 2004 {\em Phys. Rev. D\/} {\bf 70}(7) 074018

\bibitem{Metlitski2005}
Metlitski M~A and Zhitnitsky A~R 2005 {\em Phys. Rev. D\/} {\bf 72}(4) 045011

\bibitem{Abelev2009b}
Abelev B~I and et~al (STAR Collaboration) 2009 {\em Phys. Rev. Lett.\/} {\bf
  103}(25) 251601

\bibitem{Abelev2010b}
Abelev B~I and et~al (STAR Collaboration) 2010 {\em Phys. Rev. C\/} {\bf 81}(5)
  054908

\bibitem{Ajitanand2010}
Ajitanand N~N, Esumi S and Lacey R~A (PHENIX Collaboration) 2010  In Proc. of
  the RBRC Workshops (Brookhaven National Laboratory, Upton, NY, 2010), Vol.
  96.

\bibitem{Selyuzhenkov2012}
Selyuzhenkov I and Collaboration A 2012  (\textit{Preprint}
  \eprint{arXiv:1203.5230v1})

\bibitem{Burnier2011a}
Burnier Y, Kharzeev D~E, Liao J and Yee H~U 2011 {\em Phys. Rev. Lett.\/} {\bf
  107} 052303

\bibitem{Poskanzer1998}
Poskanzer A~M and Voloshin S~A 1998 {\em Phys. Rev. C\/} {\bf 58} 1671--1678

\bibitem{Dunlop2011}
Dunlop J~C, Lisa M~A and Sorensen P 2011 {\em Phys. Rev. C\/} {\bf 84}(4)
  044914

\bibitem{Xu2012a}
Xu J, Chen L~W, Ko C~M and Lin Z~W 2012 {\em Phys. Rev. C\/} {\bf 85}(4) 041901

\bibitem{Ackermann2003}
Ackermann K~H and et~al (STAR Collaboration) 2003 {\em Nucl. Instrum. Methods
  A\/} {\bf 499} 624 -- 632

\bibitem{Anderson2003}
Anderson M and et~al (STAR Collaboration) 2003 {\em Nucl. Instrum. Methods A\/}
  {\bf 499} 659 -- 678

\bibitem{Bilandzic2010}
Bilandzic A, Snellings R and Voloshin S 2011 {\em Phys. Rev. C\/} {\bf 83}
  044913

\bibitem{Gyulassy1994}
Gyulassy M and Wang X~N 1994 {\em Computer Physics Communications\/} {\bf 83}
  307 -- 331

\bibitem{Abelev2009}
Abelev B~I and et~al (STAR Collaboration) 2009 {\em Phys. Rev. C\/} {\bf 79}
  034909

\end{thebibliography}

\end{document}